Comment on "Mn Interstitial Diffusion in (Ga,Mn)As"

M. Adell[1], J. Kanski[1], L. Ilver[1], V. Stanciu[2], P. Svedlindh[2]

[1] Department of Experimental Physics, Chalmers University of Technology, SE-41296 Göteborg, Sweden

[2] Departmet of Engineering Sciences, Uppsala University, Box 534, SE-75121 Uppsala, Sweden,

In the quest for ferromagnetic semiconductors suitable for future spintronics, the (GaMn)As system remains a potential candidate material, particularly after the discovery of possibilities to raise its Curie temperature ($T_C$) by post-growth annealing [1]. Some important experimental observations concerning the annealing process are well established. The surface conditions play an important role - until now all successful annealing treatments have been performed in air or in $N_2$ atmosphere, while annealing in vacuum or with a protecting GaAs capping is found to be inefficient. The treatment also becomes gradually less efficient with increasing thickness of the (GaMn)As layer. While it is generally agreed that the annealing effects are due to diffusion and surface passivation of Mn interstitials, detailed understanding of the process is still lacking. In a recent Letter [2] Edmonds et al. assumed that Mn diffusion is the rate-limiting factor for annealing-induced changes. From measurements of layer resistivity as function of annealing time these authors then determined the activation energy for Mn diffusion. However, there is an obvious alternative rate limiting mechanism, namely the trapping efficiency of the diffusing Mn interstitials (as suggested by the mentioned surface sensitivity). In this Comment we demonstrate that this mechanism is indeed the active one in the present case, and this invalidates the results in Ref. 2.

At low Mn concentrations (in the range of a few at%) the (GaMn)As surface is essentially the same as GaAs, which is known to be quite inert. This explains why annealing of as-grown (GaMn)As layers in vacuum is inefficient. In air or in $N_2$ atmosphere the possibility to form stable Mn oxides or nitrides should of course increase the efficiency to trap the diffusing Mn atoms. Following this reasoning, we expect the Mn trapping efficiency to be even higher on a surface covered with a condensed layer of a species that forms a stable compound with Mn. In a (GaMn)As growth chamber the obvious choice is to deposit an amorphous As layer on the (GaMn)As surface prior to annealing, expecting that the diffusing Mn atoms should be bonded to As to form MnAs. With a thick amorphous As layer on the surface, annealing in air or in vacuum should give equivalent results, as verified experimentally. In this comment we only present results on samples annealed in air.

In Figure 1 we show the development of $T_C$ as function of annealing time. The $T_C$ values were obtained from SQUID magnetization measurements. All samples contained 6% Mn, and the annealing temperature was 180 °C. The first important observation is that the optimum annealing time (maximum $T_C$ corresponding to minimum resistivity) is around 2 hours. This is 1-2 orders of magnitude shorter than reported in Ref. 2, although the annealing temperature was around 10 °C lower in our case. The annealing process discussed in Ref. 2 was obviously not limited by diffusion in GaMnAs. The second important observation in Fig.1 is that the optimum annealing time is independent of layer thickness, in sharp contrast to what is reported in Ref.2 (although the final $T_C$ falls with increasing thickness as reported in other studies [1]). We ascribe this reduction of $T_C$ to the formation of a reacted layer that hinders out-diffusion of Mn interstitials. With a fraction of Mn atoms in interstitial sites of 17% [3], a full MnAs layer formed by annealing would correspond to the content of Mn interstitials in an approximately 28 nm thick $Ga_{0.94}Mn_{0.06}As$ layer. Thus, only layers thinner than this can be depleted of interstitial Mn by this process. Turning finally to the question of thickness-dependent rate of annealing, we ascribe the difference between the present results and those in Ref.2 to the formation of a surface oxide layer in the latter case. Due to the inefficient Mn trapping mechanism on the GaMnAs surface, one cannot neglect the gradual increase of the surface oxide thickness during the annealing process. Thus, with increasing oxide layer thickness, the reaction rate between the diffusing Mn and the ambient must be reduced. This effect is of course absent for annealing under an As capping layer. So, to the extent that the data in Ref. 2 reflect a diffusion process, we conclude that this is diffusion in the surface oxide layer rather than in GaMnAs.

Figure caption

Fig. 1. $T_C$ vs. annealing time for $Ga_{0.94}Mn_{0.06}As$ layers with different thickness. The annealings were carried out in air at 180 °C.

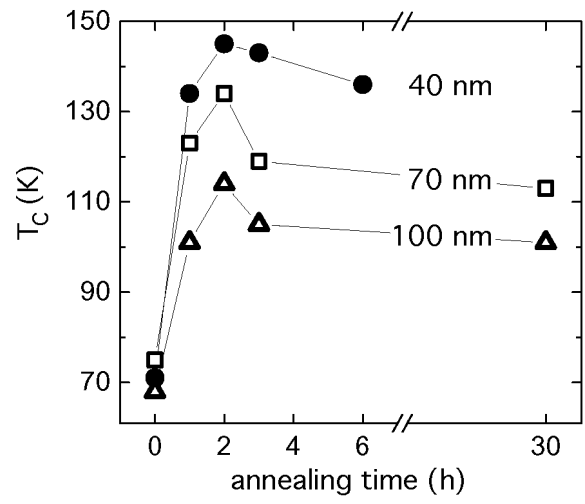